УДК 524.354.6+530.12

**A.A. Uleiev[1], A.G. Magner[1,*], S.P. Maydanyuk[2,3], A. Bonasera[4], H. Zheng[5], S.N. Fedotkin[1], A.I. Levon[6], U.V. Grygoriev[1], T. Depastas[4]**

*[1]Institute for Nuclear Research, National Academy of Sciences of Ukraine, Nuclear Theory Department, Kyiv, Ukraine*

*[2]Institute for Nuclear Research, National Academy of Sciences of Ukraine, Nuclear Processes Theory Department, Kyiv, Ukraine*

*[3]Institute of Modern Physics, Chinese Academy of Sciences, Lanzhou, China*

*[4]Cyclotron Institute, Texas A&M University, College Station, Texas, USA*

*[5]School of Physics and Information Technology, Shaanxi Normal University, Xi'an, China*

*[6]Institute for Nuclear Research, National Academy of Sciences of Ukraine, Heavy Ions Physics Department, Kyiv, Ukraine*

*Corresponding author: alexander.magner66@gmail.com

# SLOW MACROSCOPIC ROTATIONS OF NEUTRON STARS WITH A DEFORMED EFFECTIVE SURFACE

The macroscopic model for a neutron star (NS) as a finite perfect fluid at the equilibrium is extended to rotating deformed systems by incorporating the linear perturbation expansion over a small frequency $\omega$ near the Schwarzschild gravitational metric within the effective-surface (ES) approach. The NS angular momentum $I$ and moment of inertia (MI) for a slow stationary azimuthal rotation around the symmetry axis is calculated by using the Kerr metric approach in spheroidal coordinates, and compared with the spherical results. The gradient surface terms of the macroscopic NS energy density $E(\rho)$ (Equation of State) are taken into account along with the volume ones at the leading order of the leptodermic parameter $a/R \ll 1$, where $a$ is the ES crust thickness and $R$ is the NS effective radius. The analytical macroscopic NS MI expressions, $\Theta = dI/d\omega = \widetilde{\Theta}/(1-\mathcal{T}_{t\varphi})$, have been obtained in terms of the statistically averaged MI, $\widetilde{\Theta}$, and its time and azimuthal-angle $t,\varphi$ correlation, $\mathcal{T}_{t\varphi}$, as sums of the volume and surface components. The MI $\Theta$ is changed significantly as function of the effective radius $R$ and deformation parameter because of a strong gravity. We found the additional narrower constraints for the NS radius due mainly to the $t,\varphi$ correlations and surface contributions, with NS surface deformations. The adiabaticity conditions for application of the linear perturbation theory is carried out for several neutron stars with a strong gravity and relatively large rotation periods.

# I. INTRODUCTION

Tolman suggested the model for neutron stars (NSs) as a perfect finite-dense fluid under a strong gravity at zero temperature equilibrium [1, 2]. This suggestion is based on the Schwarzschild external and internal results [3–5] for a spherical metric within the General Relativity Theory (GRT). As well-known [6, 7], the NS mass M is larger or of the order of the Sun mass, $M \sim (1-3)M_{\odot}$, but the NS radius $R$ is much smaller than the Sun radius, $R \sim 10-13$ km. Therefore, the inner part $\overline{\rho}$ of the mass-baryon density $\rho$ is larger or of the order of nuclear matter density, $\overline{\rho} > \rho_0 \sim 10^{14}$ g/cm$^3$; see Fig. 1. As shown schematically in this figure, the density $\rho$ has the saturation region and continuous sharp decline within the inner crust thickness $a \sim 1$ km. Thus, there is a small leptodermic parameter, $a/R \ll 1$. Following the Tolman ideas, the Tolman-Oppenheimer-Volkoff (TOV) equations, which are very popular in nuclear astrophysics calculations for NSs, were derived by Oppenheimer and Volkoff in Ref. [9]. With the Equation of State for the same NS system, $\mathcal{E} = \mathcal{E}(\rho)$, where $\mathcal{E}(\rho)$ is the energy density, one can obtain the radial dependence of the pressure $\mathcal{P}(r)$ inside of a neutron star.

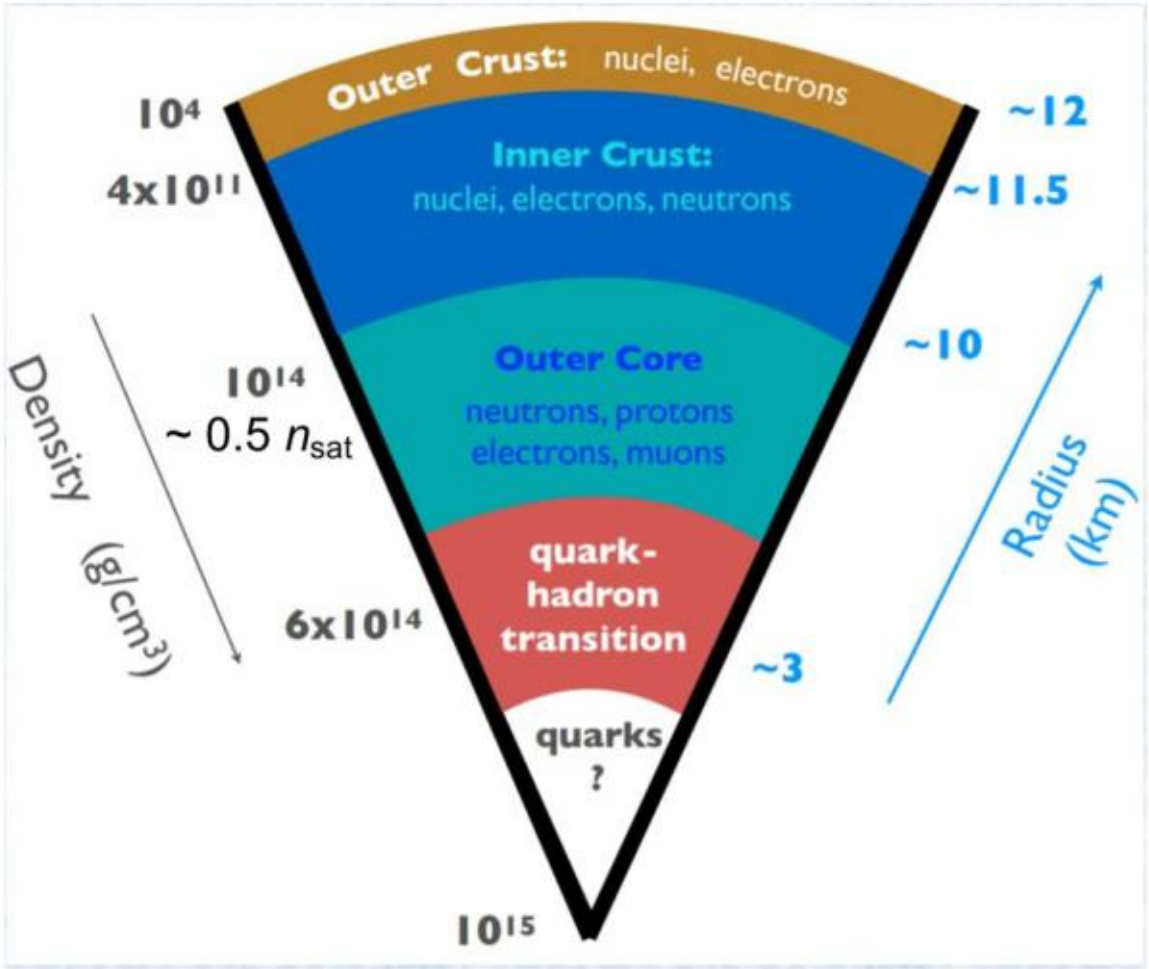

FIG. 1. Schematic typical cut of a neutron star with its proposed hadron, quark, nuclei and electron structure. Left scale: the mass-baryon density. Right one: the distance from its center (from Ref. [8]; see also Ref. [7]).

For slow NS rotations, the Kerr metric of the GRT was suggested in Ref. [10]; see its outer ($r > R$) [11] and inner ($r \leq R$) [12, 13] branches in spherical coordinates; and also Refs. [14, 15]. They both have the non-rotating spherical limit to the corresponding branch of the Schwarzschild metric [2–4]. Independently, Hartle and Thorne developed more specified theory of slow rotating deformed stars, see Refs. [16, 17]; also Refs. [18–24]. Note that the rotating star description based on the diagonal gravitational metric in spheroidal coordinates was suggested in Ref. [25]. The main question which is still open is to take into account the finite NS radius $R$, and relatively, the two (outer $r > R$ and inner $r \leq R$) branch solutions for the gravity metric of a slowely rotating axially-symmetric NS in the leading order perturbation of the Schwarzschild results which are taken at zero order [2–4].

For a good hint to the nuclear astrophysics; many astrophysical data, e.g., Refs. [26–41] on simultaneous measurements (or model calculations) of the NS masses and radii can be involved as tests for the Equation of States (EoS). Many theoretical works should be referred in this respect; see, e.g., Refs. [7, 37, 42–72]. Clear specific definitions and complete updated results for the energy density $\mathcal{E}(\rho)$ with density gradient (surface) terms and for equations of the infinite matter state with many inter-particle forces in the non-relativistic and relativistic cases can be found in the recent reviews, Refs. [61, 65]. As the TOV equations [9] are successfully used along with the EoS in a lot of astrophysical works on the NS, it would be logical to agree the arguments for the specific derivations of both these equations. This is the main motivation of our Effective Surface Approximation (ESA) and TOV macroscopic approaches [73–76]. Notice also that, of course, in nuclear astrophysics we have to take into account a strong gravity as is done in the derivations of TOV equations, unlike the traditional nuclear physics. Our suggestion for this coordination does not exclude more microscopic EoSs. However, it requires the corresponding modification of TOV equations which, unfortunately, makes the problem to be more complicate than the present analytical approach.

The gradient surface components of the energy density $\mathcal{E}(\rho)$ were suggested basically in many works; see, e.g., Refs. [7, 43, 48–51, 54, 55, 57, 62–69, 71–73, 77, 78]. Within the leptodermic approximation [73–75] based on such gradient density components of the $\mathcal{E}(\rho)$, one can consider first analytically a dense one-component system. In these calculations we take into account the energy density functional $\mathcal{E}(\rho)$ in a rather ge7neral form, as linear in $(\nabla\rho)^2$ surface terms, with the coefficients which are smooth functions of the density $\rho$. In such a system, the density $\rho$ in the leading order approximation is function of the radial coordinate with exponentially decreasing behavior from almost a constant saturation value inside the dense system to that of zero continuously through the NS effective surface in a relatively small crust range $a$. The deformed ES in the ESA is defined as the inflection points of spatial coordinates where density gradients have a sharp maximum. In order to obtain the analytical solutions for the density and EoS, we use the leptodermic macroscopic approximation. In this approximation, the gravitational field can be taken into account through the gravitational energy density part $\mathcal{U}(\rho)$, determined by its second order expansion over deflection of the density $\rho$ from its saturation value $\overline{\rho}$ (see Refs. [73–75]). Using the ESA $a/R \ll 1$, simple and accurate solutions of many nuclear and liquiddrop problems involving the density distributions were obtained for nuclei [79–84]; see also Refs. [73–75] for applications to neutron stars. The ESA exploits the property of saturation of the continuous density $\rho$ inside the system, which is a characteristic feature of dense systems such as molecular systems [85–87], liquid drops (amorphous solids), nuclei, metallic clusters [77, 78], and presumably, NSs, in line with the TOV derivations. As a result, relatively a narrow edge region exists in finite nuclei or NS (crust) in which the density drops continuously but relatively much from its almost central value to zero. We assume here that the part inside of the system far from the ES can be relatively changed a little (saturation property of a dense system).

The condition of the NS variational equilibrium means that the variation of the total energy $E$ over the density $\rho$ is zero under the constraints which fix some integrals of motion beyond the energy $E$ by the Lagrange method. The Lagrange multipliers can be determined by these constraints within the local energy-density theory, in particular, the extended Thomas-Fermi (ETF) approach from nuclear, dense molecular, and metallic-cluster physics [77, 78, 88]. The variational equation mainly determines approximately the density distribution across the diffused surface layer. The latter is of the relatively small order over the ratio of the diffuseness parameter $a$ to the (mean curvature) ES radius $R$, $a/R \ll 1$, for sufficiently heavy systems. A small parameter, $a/R$, of the expansion within the ESA can be used analytically for solving the variational problem for the minimum of the energy density with constraints for a fixed particle number, and other integrals of motion, such as angular momentum, quadrupole deformation, tidal deformations of the binary NSs, etc. Our ESA is based on the catastrophe theory for solving differential equations with a small parameter of the order of $a/R$ as the coefficient in front of the high order derivatives in the normal-to-ES direction. A relatively large change of the density $\rho$ on a small distance $a$ with respect to the curvature radius $R$ takes place for finite dense liquid-matter systems mentioned above. Within such dense systems, the density ρ is changed a little around constant saturation density $\overline{\rho}$. Therefore, including also the second order expansion of the gravitational energy density component $\mathcal{U}(\rho)$ near $\rho$, one obtains essential effects of the surface capillary pressure which were originally studied for dense molecular systems by van der Waals; see Refs. [86, 87]. The contributions of the NS surface region is negligible for the total energy and multipole moment but they are critical for local quantities as those in the energy density and pressure which are important for the system equilibrium as the capilary van der Waals effect on the curved surface.

The accuracy of the ESA was checked in Ref. [81] for the case of static and dynamic nuclear physics by comparing the results with the existing nuclear theories like Skyrme Hartree-Fock (HF) [89, 90] and ETF [78, 88]. The extension of the ES approach to the nuclear isotopic asymmetry and spin-orbit interaction has been done in Refs. [82–84]; see also Refs. [91–93]. The Swiatecki derivative terms of the symmetry energy for heavy nuclei [94–96] (see also references inside of these publications) were taken into account within the ESA in Ref. [84]. The discussions of a progress in nuclear physics and astrophysics within the relativistic local density approach can be found in review articles, Refs. [61, 65, 97]; see also Refs. [54, 66, 71, 98]. In Ref. [73], the leptodermic corrections to TOV equations were derived within the ESA for neutron stars.

In the present work we derive the coordinate dependence of a Kerr component of the gravitational metric from the GRT in the linear perturbation approximation (LPA) over the rotation frequency in terms of the two (outer $r > R$ and inner $r \leq R$) branches for the deformed axially-symmetric NS surface shapes. The corresponding NS moments of inertia are calculated as well explicitly in terms of the volume and surface components. The main focus is concentrated on the rotational constraints, additional to those obtained for nonrotating neutron stars.

## II. KERR METRICS, NS ANGULAR MOMENTUM AND EOS

The time-length element in spherical coordinates can be found for slow NS rotations around the symmetry axis from the GRT in the form ($c = G = 1$) [10]

$$ds^2 = e^{\nu} dt^2 + 2\tau\overline{\omega} M \sin^2\theta \, dt d\varphi -$$
$$-e^{\lambda} dr^2 - r^2 d\theta^2 - r^2 \sin^2\theta \, d\varphi^2, \qquad \overline{\omega} = I/M^2 \ll 1. \tag{1}$$

Within the first order LPA, $\overline{\omega} \ll 1$, $\lambda$ and $\nu$ at zero order are the Schwarzschild solution for the spherically symmetric NS surface shape (see Ref. [2])

$$e^{\lambda} = \frac{1}{1 - {r_g}/{r}}, \qquad e^{\nu} = 1 - \frac{r_g}{r}, \qquad r > R; \tag{2}$$

$$e^{\lambda} = \frac{1}{1 - {r^2}/{R_S^2}}, \qquad e^{\nu} = \left( A - \frac{1}{2}\sqrt{1 - \frac{r^2}{R_S^2}} \right)^2, \qquad r \leq R, \tag{3}$$

where $r_g$ is the gravitational radius, $R_S$ is another Schwarzschild radius determined by the inner density $\overline{\rho}$,

$$r_g = 2M, \qquad R_S = \sqrt{\frac{3}{8\pi\overline{\rho}}}, \qquad A = \frac{3}{2}\sqrt{1 - \frac{R^2}{R_S^2}}. \tag{4}$$

For the derivations of the gravitational metric $g_{\mu\nu}$ we may begin with the simplest GRT equation [2, 5]:

$$-8\pi T_{\mu}^{\nu} = R_{\mu}^{\nu} - \frac{1}{2} g_{\mu}^{\nu} \operatorname{Tr} \mathcal{R}, \tag{5}$$

where $R_{\mu}^{\nu}$ is the standard tensor Ricci, $\operatorname{Tr} \mathcal{R} = g_{\mu\nu} R^{\mu\nu}$ is the scalar invariant. The energymomentum tensor for the perfect dense finite fluid, $T_{\mu}^{\nu}$, is given by [85]

$$T_{\mu}^{\nu} = \mathcal{E}(\rho) u_{\mu} u^{\nu} + \mathcal{P} g_{\mu}^{\nu}, \tag{6}$$

where $\mathcal{P}$ is the pressure, $\mathcal{E}(\rho)$ is the energy density in the free energy ensemble, and $u_\mu$ the particle four-velocity. The GRT Eq. (5) in the first-order perturbation approach can be reduced to Eq. (A2) for $\tau(r,\theta)$ from Eq. (1) (Appendix A) and found in a separating form:

$$\tau(x,\theta) = \mathcal{X}(r)\mathcal{Y}(\theta). \tag{7}$$

As shown in Appendix A, one obtains Eq. (A3) for $\mathcal{Y}(\theta)$ which can be reduced to the equation for the hypergeometric function with the following solution:

$$\mathcal{Y}(\theta) = F\left(\alpha_+, \alpha_-; 2; \cos^2\left({}^{\theta}/_{2}\right)\right), \tag{8}$$

where the constants $\alpha_\pm$ are given by

$$\alpha_\pm = \frac{3 \pm \sqrt{9-4q}}{2}, \tag{9}$$

and $q$ is the separating constant which is the gravitational integral of motion. Notice that for $q = 0$, one finds identically $F = 1$, i.e., the symmetrical solution, Eq. (7) (see Refs. [75, 76]), for $\tau$. Therefore, the two-branch Schwarzshild solution [see Eqs. (2) and (3)] leads to the first-order LPA rotational correction for the corresponding two-branch (outer $r > R$ and inner $r \leq R$) multiplier $\mathcal{X}(r)$ of $\tau$, Eq. (7). As shown in Appendix A, for the simplest solutions for $R/R_S \ll 1$, one obtains

$$\mathcal{X}(r) = \mathcal{X}_{out} \approx \left(\frac{r}{r_g}\right)^{\frac{1-\sqrt{9-4q}}{2}}, \qquad r > R, \qquad q < 2, \tag{10}$$

and

$$\mathcal{X}(r) = \mathcal{X}_{in} \approx c_1\sqrt{\frac{r}{R_S}}J_p\left(\sqrt{5}\frac{r}{R_S}\right), \qquad r \leq R \tag{11}$$

where

$$p = \frac{1}{2}\sqrt{9-4q-6\xi^2(2-q)}, \qquad \xi = \frac{R}{R_S}. \tag{12}$$

Using the boundary condition, $\mathcal{X}_{out}(R) = \mathcal{X}_{in}(R)$ at zero order [2],

$$\frac{r_g}{R} = \frac{R^2}{R_S^2}, \tag{13}$$

one can find approximately the constant $c_1$,

$$c_1 = \frac{\xi^{-2\left(1-\sqrt{9-4q}\right)-1/2}}{2J_p\left(\sqrt{5}\xi\right)}. \tag{14}$$

For the spherical case $q = 0$ and $r > R$, one obtains from Eq. (10) the well-known Boyer-Lindquist result [5, 11] for vacuum (mass-point solution): $\mathcal{X}_{out}(r) = r_g/r$. Generally speaking, the separation constant $q$ is the gravitational integral of motion for a given $\xi$ beyond the total energy $E$ and azimuthal angular momentum $I$ introduced above. Notice also that our result, Eqs. (8), (10) and (11), differ from those found in Ref. [18] because they were derived for a finite NS radius $R$ in terms of the outer-inner branches, in line of those of the Schwarzschild solution [3, 4]. We use the opportunity for separation of variables $r$ and $\theta$ in Eq. (A2) for $\tau(r,\theta)$, instead of the multipole expansion over spherical harmonics in vaccum medium (mass-point) approximation.

For the angular momentum $I$ in the case of a slow stationary rotation of the perfect finite fluid around its symmetry axis, one has the following expression [5, 20–24, 85]:

$$I = \int T_\mu^\nu \phi^\mu \hat{n}_\nu \, d\mathcal{V}, \tag{15}$$

where $T_\mu^\nu$ is the energy-momentum tensor, Eq. (6), $\mu,\nu = \{0,1,2,3\}$, and approximately, $I = \omega\Theta$, with the NS moment of inertia $\Theta$. Within the considered first-order perturbation approach, the integration is carried out over the whole three dimensional volume accounting for the gravity schortenning, $d\mathcal{V} = \mathcal{J}(r)\mathrm{d}r \sin\theta \, \mathrm{d}\theta \mathrm{d}\varphi$, where $\mathcal{J}(r)$ is the volume transformation Jacobian,

$$\mathcal{J}(r) = \frac{1}{\sqrt{1 - r^2/R_S^2}}, \qquad r \leq R \tag{16}$$

$$= \frac{1}{\sqrt{1 - r_g/r}}, \qquad r > R. \tag{17}$$

In Eq. (15), the energy-momentum tensor $T_\mu^\nu$ in the canonical ensemble for the rotating perfect fluid [85] was expressed in terms of the energy density $\mathcal{E}(\rho)$, pressure $\mathcal{P}$, and fourvelocity $u^\mu$; see Eq. (6). The ETF EoS is taken in a rather general macroscopic statistically averaged form as sum of the smooth volume and gradient surface terms (Appendix B),

$$\mathcal{E}(\rho) = \rho + \varepsilon_G(\rho) + \mathcal{C}(\nabla\rho)^2, \tag{18}$$

where

$$\varepsilon_G(\rho) \propto K_G \rho(\rho - \overline{\rho})/\overline{\rho}^2, \qquad K_G = K_N + 9\overline{\rho}^2 \mathcal{U}_G''(\overline{\rho}). \tag{19}$$

Here, $\mathcal{U}_G''$ is the second derivative of the gravitational part $U_G(\rho)$ of the energy density $\mathcal{E}(\rho)$ in the leptodermic expansion,

$$\mathcal{U}_G(\rho) = \mathcal{U}_0 + \frac{1}{2}\mathcal{U}_2(\rho - \overline{\rho})^2. \tag{20}$$

## III. NS MOMENTS OF INERTIA AND ROTATION CONSTRAINTS

The NS moment of inertia (MI) $\Theta$ can be derived from Eqs. (15) and (6) in terms of the statistically averaged $\widetilde{\Theta}$ and correlations $\mathcal{T}_{t\varphi}$ contributions,

$$\Theta = \frac{\partial I}{\partial \omega} = \frac{\widetilde{\Theta}}{1 - \mathcal{T}_{t\varphi}}, \tag{21}$$

where

$$\widetilde{\Theta} = \int \mathcal{E}(\rho) e^{-\nu} r^2 \sin\theta \, \mathrm{d}\mathcal{V} = \widetilde{\Theta}_V + \widetilde{\Theta}_S, \tag{22}$$

$$\mathcal{T}_{t\varphi} = \frac{2}{M} \int \mathcal{E}(\rho) e^{-\nu} \tau \sin\theta \, \mathrm{d}\mathcal{V} = \mathcal{T}_V + \mathcal{T}_S. \tag{23}$$

The correlation function $\tau(r,\theta)$ is the GRT solution (7) found in a separation form within the first order perturbation theory over a small parameter $\omega = I/M^2$ (Appendix A). Using the expression (18) for the energy density $\mathcal{E}(\rho)$, one approximately obtains the very right volume and surface decomposition which is similar to that for the total energy $E$ (Appendix B). The MI expression (21) for $\Theta$ has a pole structure. In order to prove that we substitute Eq. (6) for the energy-momentum tensor of the ideal dense fluid [85] into Eq. (15) for the angular momentum $I$. For a fourth velocity $u_\mu$ of the particle motion, one has

$$u_\mu = u^t \big(\omega \phi_\mu + t_\mu\big), \qquad u^t = \frac{e^{-\nu/2}}{(1 - v^2)^{1/2}} \approx e^{-\nu/2}, \tag{24}$$

where $\phi_\mu$ and $t_\mu$ are Killing vectors which determine a slow stationary rotation with a constant frequency $\omega$ around the azimuthal axis under a gravitational field [20–23]. The first component in brackets, $\omega\phi_\mu$, is related to the global rotation as a whole, like for a rigid body. The second part, $t_\mu$, is proportional to the off-diagonal correlation element of the gravitational metric, $g_{t\varphi} \propto \omega$. The particle velocity $u_\mu$ is normalized to one, $u_\mu u^\mu = 1$, and $u^t \approx e^{-\nu/2}$ for relatively slow particle velocities $v/c$. With Eq. (24) and the definition of the Killing vectors (see, e.g., Refs. [21, 23, 75]), one obtains the corresponding decomposition for the angular momentum $I$:

$$I = \int \mathcal{E}(\rho) (u^t)^2 \big(\omega g_{\varphi\varphi} + g_{t\varphi}\big) \mathrm{d}\mathcal{V}, \tag{25}$$

where $g_{t\varphi} \propto \omega\tau\Theta$, Eq. (1), and $\tau$ is the solution of the GRT equation, Eqs. (7), (8) and (10) with (11). Differentiating now over the frequency $\omega$, with the definition of the MI [see Eq. (21)], one may express the MI $\Theta$ through the corresponding "rigid-body"-like and the gravity correlation $\mathcal{T}_{t\varphi}$ components,

$$\Theta = \frac{\partial I}{\partial \omega} \int \mathcal{E}(\rho) e^{-\nu} r_\perp^2 \left(1 + \frac{2\tau}{r^2} \frac{\Theta}{M}\right) \mathrm{d}\mathcal{V}. \tag{26}$$

Solving this simple algebraic self-consistency relation with respect to MI $\Theta$, one arrives at the pole expression (21). Notice, this pole structure can be proved within the perturbation theory which includes also the $\omega^2$ corrections; see Refs. [11, 12, 16, 17]. However, the specific expressions for the statistically averaged, $\widetilde{\Theta}$, and correlation contributions become more complicate than those in Eqs. (22) and (23).

For small frequencies of the macroscopic NS rotation, one should expect mostly an adiabatic motion. In our ESA, it obeys the condition of smallness of the rotation energy, Erot, with respect to the ETF energy $E$,

$$E_{rot} = \frac{1}{2}\Theta\omega^2 \ll E. \tag{27}$$

This condition was studied in details in Refs. [75, 76] for Hogan's and our spherical offdiagonal gravitational metric coefficient $\tau(r)$. For comparison with the observation results for the NS rotation periods $P = 2\pi/\omega$, one can conveniently specify this condition as

$$P \gg P_0, \tag{28}$$

where $P_0$ is the characteristic upper period limit for adiabatic motion,

$$P_0 = 2\pi\sqrt{\frac{\Theta}{2E}}, \tag{29}$$

$\Theta$ is the full NS MI, Eq. (21), and $E = E_{ETF}$ is the complete macroscopic energy in the non-rotating ETF neutron-star approximation, Eq. (B1); see Appendix B.

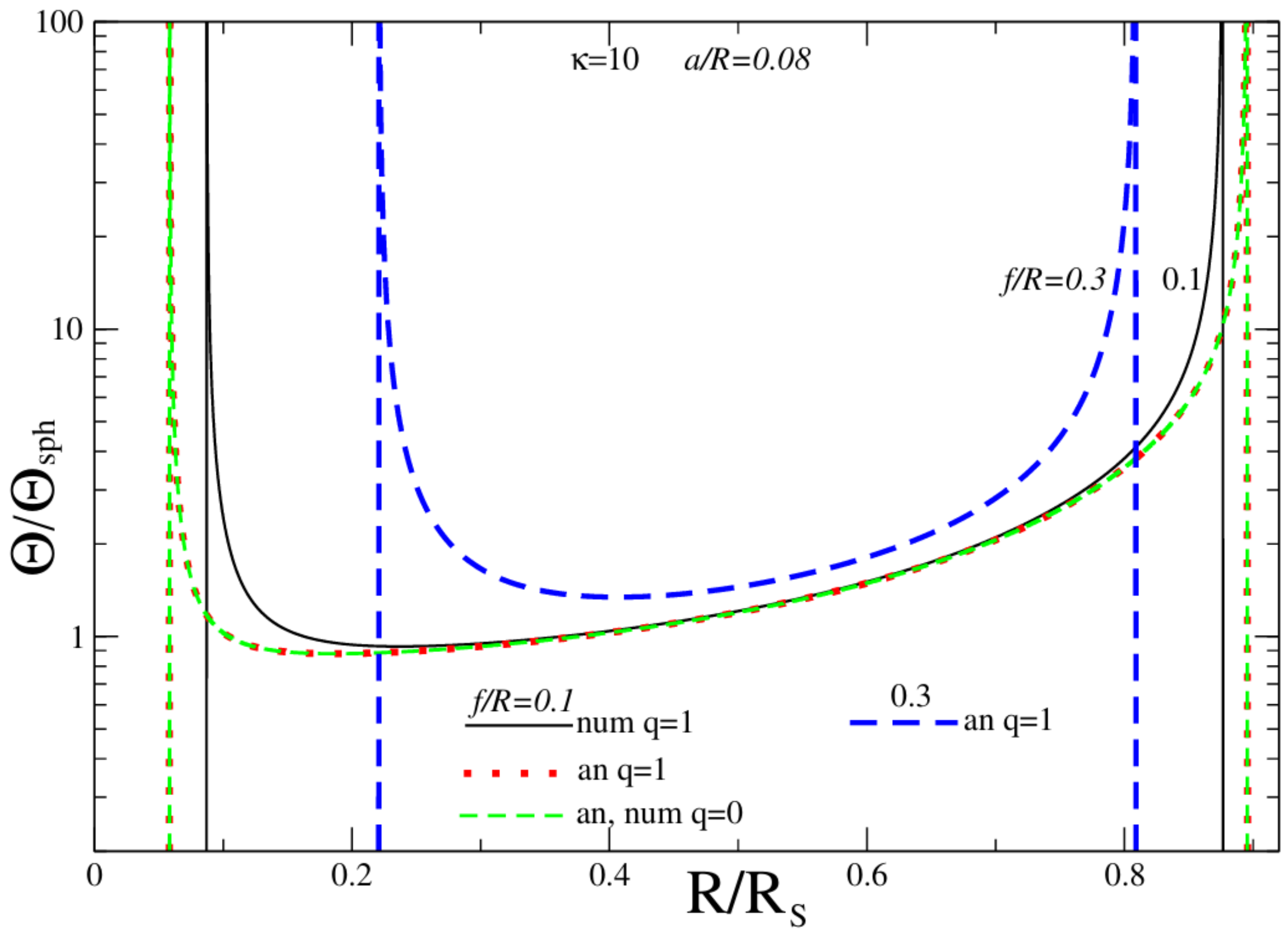

FIG. 2. The total moments of inertia $\Theta$, Eqs. (21)-(23) (in units of $\Theta_{sph} = 2MR^2/5$ ) are shown as function of the radius (in units of $R_S$, Eq. (4)). The numerical ["num", black solid line, Eq. (C4)] and approximately analytical ["an", red dotted line, Eq. (C5)] for the volume correlation contribution $\mathcal{T}_V$ and full analytical statistically averaged $\widetilde{\Theta}$ [Eq. (C2)] are presented, respectively. The relative spheroidal-focus distance for the deformed surface (deformation parameter) is $f/R = \sqrt{a^2 - b^2}/R = 0.1$ for all curves, except for the blue dashed curve where $f/R = 0.3$ ($a$ and $b$ are semi-axixes of the oblate spheroidal cavity). The crust thickness in the same units is $a/R = 0.08$, and the dimensionless nuclear-gravity incompressibility, Eq. (B5), related to a strong gravity, are given by $a/R = 0.08$ and $\kappa = 10$, respectively. The separation constant everywhere is $q = 1$ for the deformed NS surface, besides of the green dashed line which shows the results for the spherically symmetrical NS surface shape at $q = 0$. The vertical lines for any set of parameters are asymptotes related to the MI poles which present the two (lower and upper) rotational constraints boundaries $R_l$ and $R_u$, $R_l < R < R_u$ in units of $R_S$ ; see text.

## IV. DISCUSSIONS OF THE RESULTS

Figures 2-5 show our results. The moments of inertia $\Theta$, Eq. (21), as functions of the NS radius variable R in units of the Schwarzschild radius $R_S$, Eq. (4), are shown in Figs. 2 and 3 for different approaches and parameters. Figure 2 displays MI with the spheroidal-surface deformation parameter $f/R$ at the separation constant $q = 1$. As seen from this figure, one finds good agreement between the numerical ("num" black solid, with the correlation contribution $\mathcal{T}_{t\varphi}^{(V)}$, Eq. (C3)), and analytical ("an" red dotted lines, Eq. (C5)) at $f/R = 0.1$, besides of narrow ranges near the asymptotes (vertical lines). Other analytical volume statistically averaged contributions $\widetilde{\Theta}_V$, Eqs. (C1) and (C2), and both surface components $\widetilde{\Theta}_V$ and $\mathcal{T}_S$, Eqs. (C6) and (C7), are used in all these calculations because they agreed well with the corresponding numerical results. The asymptotes mentioned above arise at small ($R = R_l$) and large ($R = R_u$) radii because of the MI poles, Eq. (21). Therefore, similarly as in the Schwarzschild metric, the rotational radius constraints between these two asymptotes appear due to the surface deformation $f/R$ and $t\varphi$-correlational gravitational contribution $\mathcal{T}_{t\varphi}$, $R_l < R < R_u$. The rotation and deformation effects lead to a narrow NS radius constraint. Notice that one obtains the rotational constraint from small and large sides, in contrast to the spherical case considered in Ref. [75] where we found only the one-side, $R < R_u$, constraint. For a small surface deformation $f/R = 0.1$, the analytical expression for the MI at $q = 1$ (red dotted) is close to the spherical case $q = 0$ (green dashed lines) for the same other parameters. With increasing deformation, $f/R$ and the same dimensionless incompressibility $\kappa = 10$ [Eq. (B5)] the constraint ranges decrease much with the larger $R_l$ and smaller $R_u$ boundary radii.

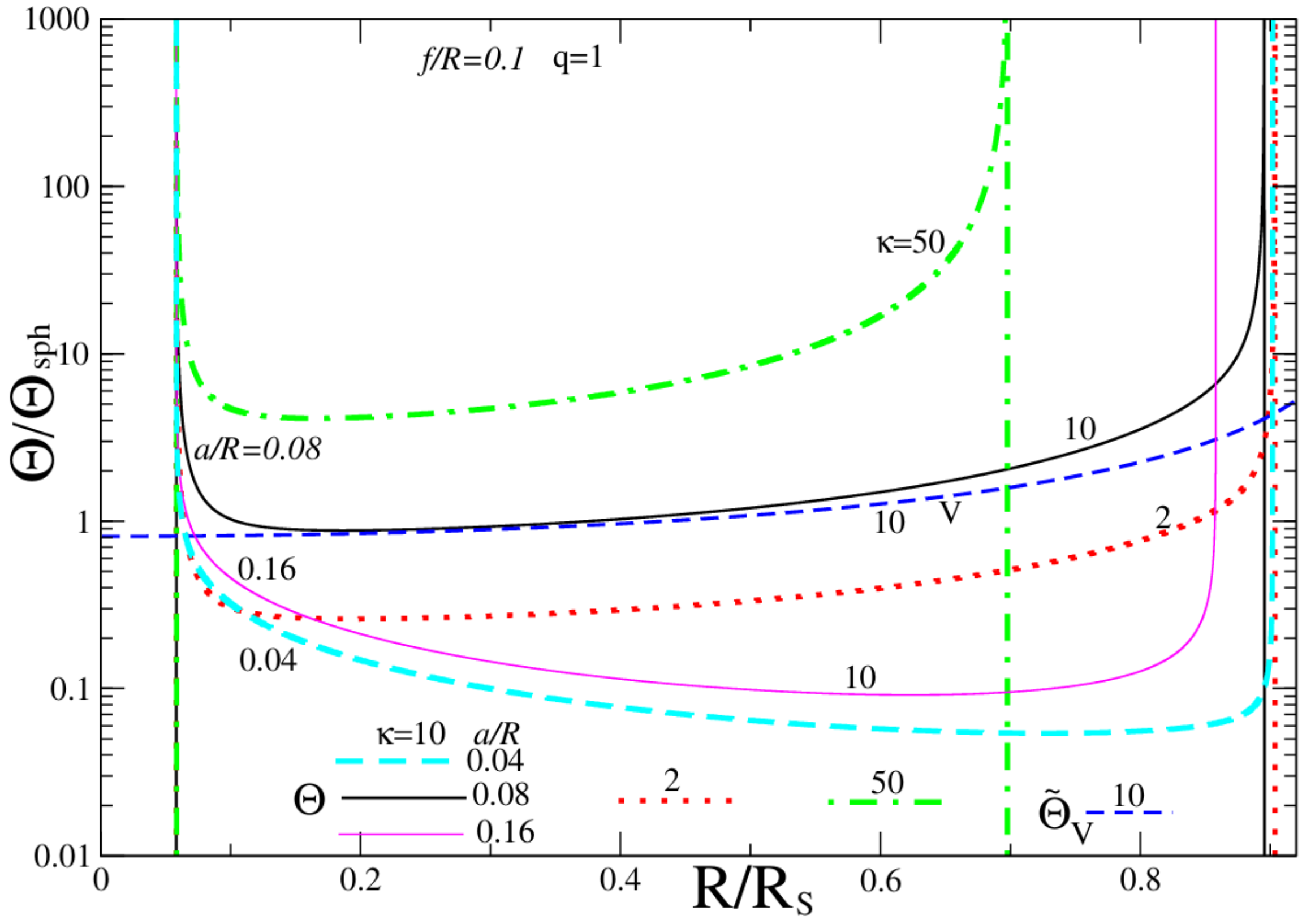

FIG. 3. The same as in Fig. 2 but that depending on the incompressibility $\kappa = 10$ (strong), Eq. (B5), 2 (weak), and 50 (super strong gravity) for the deformation $f/R = 0.1$. We show also the NS MI as function of the leptodermic parameter $a/R = 0.04$, 0.08 and 0.16. The blue dashed curve shows the statistically averaged volume MI component $\widetilde{\Theta}_V$. Vertical lines show asymptotes related to the boundaries of rotation constraints.

Figure 3 shows the incompressibility $\kappa$ and leptodermicity $a/R$ dependence of the MI as functions of the NS radius. With increasing $\kappa$, Eq. (B5) from $\kappa = 2$ (weak), through 10 (strong), to 50 (super strong gravity), the MI increases. The upper radius constraint ($R = R_u$) is significantly shifted to smaller radii $R/R_S$. It depends strongly on the gravity through the incompressibility $\kappa$ due to the surface MI contributions while the lower value ($R = R_l$) is almost independent of $\kappa$. More complicate dependence of the deformed NS MI $\Theta$ on the leptodermic parameter $a/R$, Eq. (B4), takes place. Significant combined correlation and surface effects are seen from the comparison of these dramatically changed curves with a minimum between asymptotes for the MI $\Theta$ and a smooth monotonic blue-dashed line behavior for the volume statistically averaged contribution $\widetilde{\Theta}_V$.

Notice that this lower constraint is usually much larger than the gravitational radius $r_g$, Eq. (4), which determines the event horizon [5] in the non-rotating systems. According to the boundary condition, Eq. (13), the variable value $r_g/R_S$ is much smaller than all MI poles for $R/R_S < 0.6$.

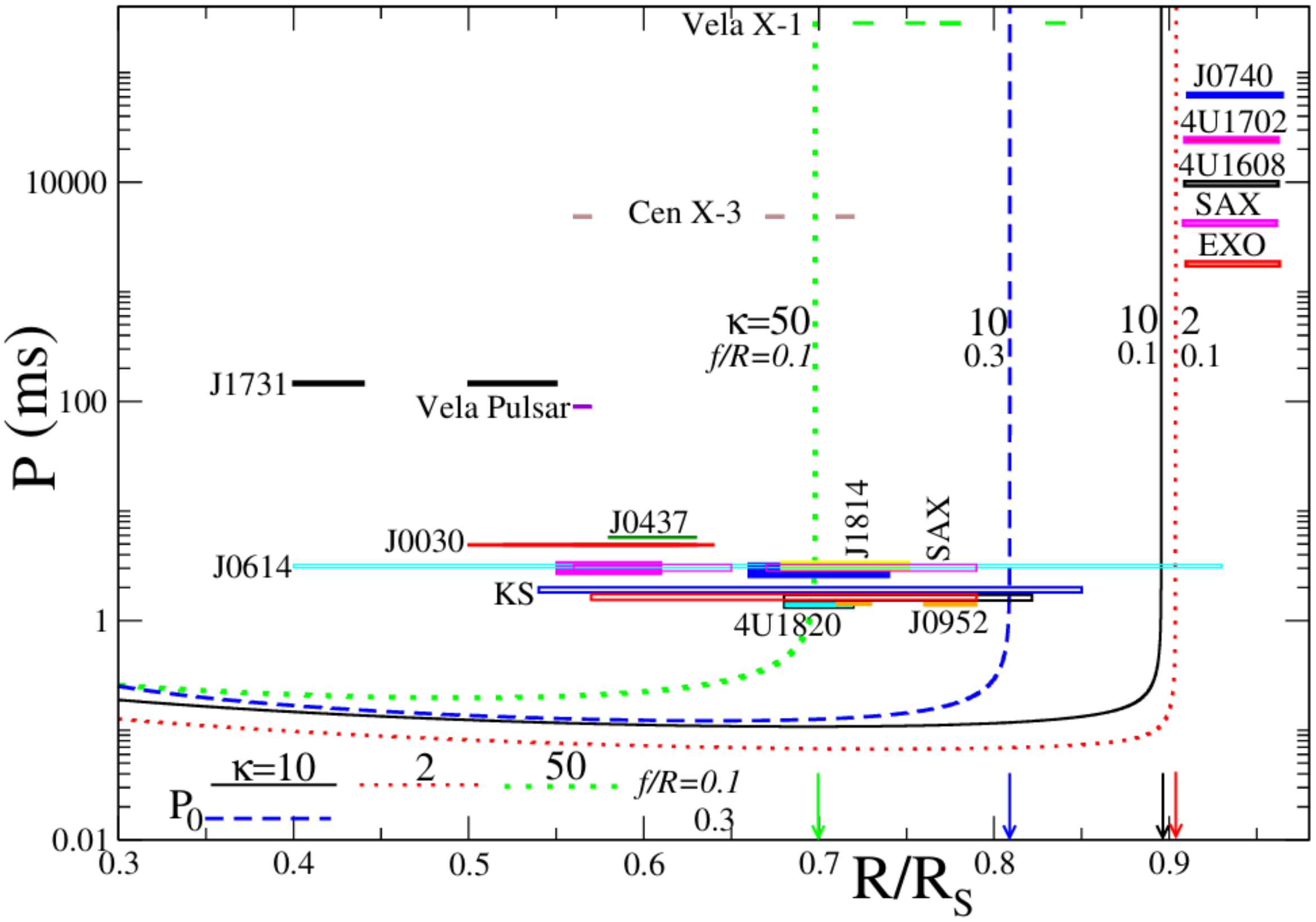

FIG. 4. NS periods $P$ (in units of ms) are plotted as functions of radius, $R/R_S$ (in units of RS, Eq. (4)). Straight line segments show different observational and model calculated data for known NSs (from Ref. [75]). Solid, dotted and dashed lines are the characteristic periods $P_0$, Eq. (28), for the full gravitational metric with accounting numerically for the $t, \varphi$ correlations and surface contributions at the relative incompressibility values $\kappa = 2$ (weak, red dashed), 10 (strong, black solid with the deformation $f/R = 0.1$ and blue dashed with $f/R = 0.3$) gravitational contribution through $\kappa$) and 50 (super strong gravity, green dotted curve), correspondingly; see Eq. (B5) for $\kappa$ and text. The leptodermic parameter $a/R = 0.08$ is the same as in other figures. The effective radius value is mean for the shown NSs, $R = 12$ km, $\xi = R/RS = [2GM/(Rc^2)]^{1/2}$; see Eq. (13) and Ref. [75].

Figure 4 displays the characteristic limit periods $P_0$, Eqs. (28) and (29), as function of the radius, $R/R_S$, in units of the Schwarzschild radius $R_S$, Eq. (4), for different parameters $f/R$, and $\kappa$ versus the results calculated for the data on the mass $M$, radius $R$, and period $P$ for several NSs; see Ref. [75]. The intervals of the variable $R/R_S$ are presented by line segments within the errors for the mass $M$ and radii $R$. For our evaluations of $P_0$, Eq. (29), the variable $R/R_S$ can be expressed through the mass $M$ and radius $R$ by using both the boundary condition for slitching the outer and inner solutions, Eq. (13), and the gravitational radius $r_g$ value, Eq. (4) (see the caption to Fig. 4). The limit periods are dramatic functions of the relative effective NS radius, $R/R_S$, for a given mean $R$, with the corresponding asymptotes shown by the arrows. The periods $P_0$ as functions of $R/R_S$ are lines convergent to the corresponding asymptotes. In Fig. 4 we also show the essential dependence of the full limit periods $P_0$ on the dimensionless incompressibility parameter $\kappa$ [Eq. (B5)] through the tension coefficient $\sigma$, Eq. (B3), in the surface energy contribution ES. The values of $P_0$ increase with $\kappa$. As a result, for a strong $\kappa = 10$ and small surface deformation $f/R = 0.1$, the adiabatic condition, Eq. (28), is carried out well for all neutron stars shown in Fig. 4, except for the NS J0614. For a weak gravitation ($\kappa = 2$), this adiabaticity condition for the same deformation becomes even better, in contrast to more approximate inner solutions discussed in Ref. [75]. With increasing the incompressibility $\kappa$, one finds worse applicability of the adiabatic condition, Eq. (28). The surface effect is enhanced because of the surface part of the correlation contribution $\mathcal{T}_{t\varphi}$, due mainly to a change of the rotational root of equation $\mathcal{T}_{t\varphi} = 1$ . Contour plots in Fig. 5 show the analytical calculations of the rotational parameter, $\omega(R/R_S, P)$, Eq. (1), depending on the dimensionless variable $\xi = R/R_S$ and rotational period $P$. The intervals mostly overlap data presented in Fig. 4, except for Cen X-3 and Vela X-1 with extremely large periods. As seen from contour plots (a,b), for the deformations $f/R = 0.1$ and incompressibilities $\kappa = 10$ (a) and $\kappa = 50$ (b) the parameter $\omega$ has a minimum but in somewhat different areas of the variable $\xi$ and with different magnitude of $\omega(R/R_S, P)$. With increasing incompressibilities $\kappa$ or deformation $f/R$, the rotational parameter values $I/M^2$ are increased for any small deformation. In both cases, the non physical areas become larger for smaller periods, in line of the results presented in Fig. 4. The dimensionless angular momenta $Ic/(M^2G)$ are mainly close for the deformed $q = 1$ to spherical $q = 0$ cases for a small deformation $f/R = 0.1$ and a strong incompressibility $\kappa = 10$. They are mostly small for the NSs J0030, J1731-347, Centaurus X3, Vela Pulsar, Vela X-1, J1814-338, and J0437-4715. For all other NSs shown in Fig. 4, one finds larger values of $Ic/(M^2G)$, which are sometimes of the order of one, or even negative. The latter is met for non-physical cases when the effective radius $R$ occurs out the constraint boundaries, as explained above. Notice that for larger deformations ($f/R = 0.3$), the situation becomes worse concerning the adiabaticity and slow rotation conditions. We would conclude that for stronger deformations, we should go beyond the leading first-order perturbation approxima tion over the parameter $Ic/(M^2G)$, accounting for quadratic and high order corrections over $Ic/(M^2G)$, or non-linear corrections in expansion over deformation parameter $f/R$.

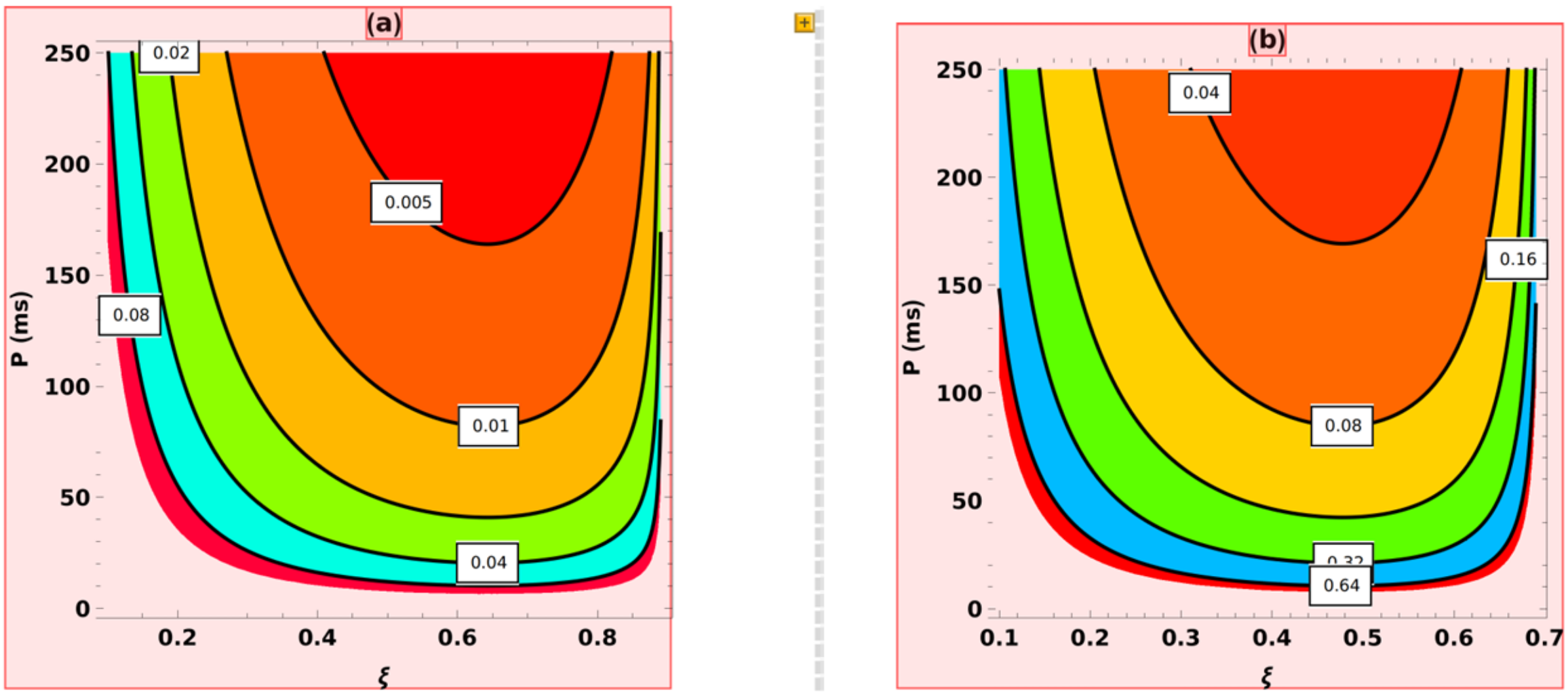

FIG. 5. Contour plots show the dimensionless NS angular momentum, $Ic/(M^2G) \approx 2\pi\Theta c/(M^2PG)$, Eq. (1), as function of the variable $\xi = R/R_S$ and period (in milliseconds) for the metric coefficient $\tau(r,\theta)$, Eq. (7) at the deformation $f/R = 0.1$ and the separation constant $q = 1$. The typical values of $Ic/(M^2G)$ are shown in squares. (a): For incompressibility $\kappa = 10$; and (b): $\kappa = 50$. A mean radius is $R = 12$ km, and a crust thickness $a/R = 0.08$, as in Fig. 4. White areas are related to non-physical parameters values outside the constraints.

Thus, as clearly seen from Figs. 4 and 5, the adiabaticity condition (28) is mainly carried out for the NSs, especially well for the NSs J1731-347, Cen X-3, Vela Pulsar, Vela X-1, and J0437-4715 at a strong gravity $\kappa = 10$ and small surface deformation $f/R = 0.1$, and becomes worse for larger incompressibility $\kappa$. In line with the results for the dimensionless angular momentum $Ic/(M^2G)$, it would be nice to take into account non-adiabatic effects related' to quadratic frequencies $\sim\overline{\omega}^2$, for several NSs, especially for the rotation of NSs J0740 and J0952. In this non-adiabatic approach, the NS MI $\Theta(\omega)$ depends essentially on $\omega$.

## V. CONCLUSIONS

The macroscopic effective-surface approximation based on the leptodermic expansion over a small parameter- a ratio of the crust thickness $a$ to the effective radius $R$ of the system- is extended for the description of neutron star rotations by taking into account the surface deformations. The neutron star was considered as a finite perfect dense-liquid (including amorphous solid) system rotating stably with a small constant frequency $\omega$ around a symmetry axis within the Kerr approach by using the spherical coordinates. Within the linear perturbation theory over small parameter $Ic/(M^2G)$ with respect to the Schwarzschild spherically symmetric system at zero order, one can approximately derive the solution to the GRT equations, which are linear in the Riemann tensor, in terms of the separable radial $r$ and polar angle $\theta$ variables. This is the two-branch solution for the radial variable $r$ in the outer $r > R$ and inner $r \leq R$ space ranges. In this approximation for deformed surfaces of neutron stars we show that the outer solution ($r > R$) for the off-diagonal gravitational metric element $g_{03}$ in the spherically symmetric limit $q = 0$ is the Boyer-Lindquist result obtained earlier for a vacuum (or for a mass-point gravitational system). But it differs essentially for the deformed NS surface at the nonzero separation constant, $0 < q < 2$. For any non-zero separation variable value $q$ for small deformations, one obtains this well known result [5] asymptotically at large distances from a neutron star. Using the macroscopic Equation of State (EoS) with the energy density, $\mathcal{E} = \mathcal{E}(\rho)$ as functional of the baryon density $\rho$, in a rather general form one can analytically take into account the gradient surface terms which are important for local NS characteristic quantities as the pressure. With the help of our approximate GRT solutions we obtained the NS adiabatic MI $\Theta$ in terms of the statistically averaged $\widetilde{\Theta}$ and rotational correlational $\mathcal{T}_{t\varphi}$ contributions. Each of them is presented as a sum of the volume and surface terms. These MIs of the NSs with the deformed effective surface change dramatically as functions of the dimensionless angular momentum $Ic/(M^2G)$ (through the off-diagonal $g_{03}$ gravitational-metric component), the effective radius $R$, and dimensionless incompressibility $\kappa$, due to the $t, \varphi$ correlational contribution $\mathcal{T}_{t\varphi}$. The significant macroscopic surface effects were also found and studied in relation to this correlation term for a strong gravity. We found the rotational constraints to the effective mean radius $R$ for slightly deformed surfaces of neutron stars, in addition to those of coming from the Schwarzschild solution.

## ACKNOWLEDGMENTS

The authors greatly acknowledge O.Y. Dzyublik, R.B. Granyk, V.Z. Goldberg, M.I. Gorenstein, V.P. Gusynin, F.A. Ivanyuk, J. Holt, C.M. Ko, V.V. Kobychev, E.I. Koshchiy, J.B. Natowitz, S.A. Omelchenko, A.I. Sanzhur, B.I. Sidorenko, P. Ring, G.V. Rogachev, Y.V. Shtanov, Yu.V. Taistra, V.I. Zhdanov for many creative and useful discussions.

## Appendix A

### Gravitational metric for deformed rotating NSs

For calculations of the correlation contribution $\mathcal{T}_{t\varphi}$, Eq. (23), one has to calculate in $ds^2$ [Eq. (1)] the $\tau(r,\theta)$ factor which is a function of the radial and polar-angle variables [10]. Taking into account the deformation of the gravitational field through $g_{03}(r,\theta)$, due to the NS rotation within the LPA on the basis of the same zero-order two-branch Schwarzschild metric solution, Eqs. (2) and (3), and assuming the azimuthal $\varphi$ invariance of the GRT solutions we will look for the gravitational field $g(r,\theta)$ in the following form [10]:

$$g_{00} = e^{\nu}, g_{11} = -e^{\lambda}, g_{22} = -r^2, g_{33} = -r^2 \sin^2\theta,$$
$$g_{03} = g_{30} = 2\tau(r,\theta)\overline{\omega} M \sin^2\theta\,, \qquad \text{(A1)}$$

and $g_{\mu\nu}$ for all other $\mu$ and $\nu$ subscripts are zeros. Here, $\nu(r)$ and $\lambda(r)$ are known Schwarzschild gravitational metric functions [see Eqs. (2) and (3)] at the zero order LPA ($\omega = 0$), $\omega$ is a small LPA parameter[1], Eq. (1), and $\tau(r,\theta)$ is unknown function of the radial $r$ and polar angle $\theta$ coordinates in the spherical coordinate system $t, r, \theta, \varphi$. Calculating also the Christoffel and Ricci tensor, one obtains from Eq. (5) the linear partial-derivative equation over $\tau(r,\theta)$ by the second order:

$$\sin\theta\,\tau(x,\theta)[2x^2\nu'' + x^2(\nu')^2 - x\lambda'(4 + x\nu') + 4] - 6e^{\lambda}\cos\theta\frac{\partial^2\tau(x,\theta)}{\partial\theta^2} -$$
$$-x^2(\lambda' + \nu')\frac{\partial\tau(x,\theta)}{\partial x} - 2\frac{\partial^2\tau(x,\theta)}{\partial x^2}] = 0, \qquad \text{(A2)}$$

where $x = r/R_S$ for $r \le R$, or $x = r/r_g$ for $r \ge R$ [see Eq. (4) for the gravitational radius rg and Schwarzschild radius RS], and $\tau(r,\theta)$ is identically re-written, for simplicity, as $\tau(x,\theta)$ in terms of the dimensionless variable $x$. We will look for the solutions in the separable way, Eq. (7). Then, the variables in Eq. (A2) are separated such that one obtains two linear ordinary equations for X(x) and Y(θ) of the second order.

**1. Angle solutions**. For $Y(\theta)$ one finds a general ordinary linear equation of the second kind, common for outer ($r > R$) and inner ($r < R$) regions:

$$\mathcal{Y}''(\theta) + 3\cot\theta\,\mathcal{Y}'(\theta) - q\mathcal{Y}(\theta) = 0, \qquad \text{(A3)}$$

where q is a separating variable constant. Changing variables, $\psi = \theta + \pi/2$, and $w = (1+u)/2$, one obtains the hypergeometric (Gauss) equation

$$w(w-1)F''_{ww} + [(\alpha + \beta + 1)w - \gamma]F'_{ww} + \alpha\beta F = 0, \qquad \text{(A4)}$$

[1] Kerr introduced another small parameter $a = \overline{\omega}M = I/M$.

which has the regular analytical solution if γ is not a negative integer,

$$\mathcal{Y}(\theta) = C_1 F(\alpha_+, \alpha_-, \gamma, w), \qquad w = \cos^2(\theta/2). \tag{A5}$$

In Eqs. (A4) and (A5), $\alpha_\pm$ are the roots of the quadratic equation $u^2 - 3u + q = 0$, and $\gamma = 2$, see Eq. (9).

**2. Outer radial solutions ($r > R$).** The equation for a radial solution $\mathcal{X}(x)$ ($x = r/r_g$) in the outer case, $r > R$, according to Eq. (2), can be simplified because of

$$a_2(x)\mathcal{X}''(x) + a_0(x,q)\mathcal{X}(x) = 0, \tag{A6}$$

where

$$a_2(x) = x^2 \exp[-\lambda(x)], \qquad a_0(x,q) = 2x^2\lambda''(x) - \\ -x\lambda'(x)[x - 4 - x\lambda'(x)] + 4 + q. \tag{A7}$$

Thus, for $X(x)$ at $r > R$, one obtains the linear ordinary second-order equation:

$$x^2(x-1)\mathcal{X}''(x) + (2 - 2x + xq)\mathcal{X}(x) = 0. \tag{A8}$$

For the limit $q \to 0$ one has

$$x^2\mathcal{X}''(x) - 2\mathcal{X}(x) = 0. \tag{A9}$$

For arbitrary $q$ constants, the solutions of Eq. (A8) look rather complicated. However, according to the boundary condition, Eq. (13), for the case of $R \ll R_S$, for which one finds simple analytical inner solutions for the $\mathcal{X}(x)$, we need only asymptotic outer solutions for large $x = r/r_g$. They can be obtained analytically in a close simple form; see the simplest solution in Eq. (10).

These asymptotic outer solutions for $x = r/r_g \gg 1$ can beobtained by using the poly nomial expansion over $1/x$ up to a definite order $k$ analytically by the standard method of independent coefficients. Up to the first order $k = 1$, one obtains the following convergent result X1(x) for the outer solution:

$$\mathcal{X}_1(x) = x^{(1-\sqrt{9-4q})/2}\left[1 + \frac{q(1-\sqrt{9-4q})}{4(2-q)x} + \mathcal{O}\left(\frac{1}{x^2}\right)\right], \qquad q < 2. \tag{A10}$$

For the leading term, one results in Eq. (10). For $q \geq 2$, one finds non-physical constant or divergent solutions in the limit $x \to \infty$.

**3. Inner radial solutions ($r \leq R$).** Substituting Eq. (7) into Eq. (A2) and using Eq. (A3) for the angle function $Y(\theta)$ and Eq. (3) for the inner $\lambda$ and $\nu$ one obtains the equation for $X(x)(x = r/R_S)$ in the inner region, $r < R$,

$$a_2(x)\mathcal{X}''(x) + a_1(x)\,\mathcal{X}'(x) + a_0(x,q)\mathcal{X}(x) = 0, \tag{A11}$$

where

$$a_0 = 2\left[6\sqrt{1-\xi^2} - (10-9\xi^2)\sqrt{1-x^2} + x^4\left(21\sqrt{1-\xi^2} - 4\sqrt{1-x^2}\right)\right.$$
$$- x^2\left(27\sqrt{1-\xi^2} - (32-27\xi^2)\sqrt{1-x^2}\right)$$
$$\left. + q\left(-6\sqrt{1-\xi^2} + (10-9\xi^2)\sqrt{1-x^2} + x^2\left(6\sqrt{1-\xi^2} - \sqrt{1-x^2}\right)\right)\right],$$
$$a_1 = 3x^3[\sqrt{1-\xi^2}(1-x^2) - 3(1-\xi^2)\sqrt{1-x^2},$$
$$a_2 = x^2(x^2-1)[6\sqrt{1-\xi^2} - (10-9\xi^2)\times\sqrt{1-x^2}$$
$$\left. - x^2\left(6\sqrt{1-x^2} - x^2\left(6\sqrt{1-\xi^2} - \sqrt{1-x^2}\right)\right)\right];$$

see Eq. (12) for $\xi$. In order to simplify Eq. (A11) for analytical solutions we expand the coefficients, Eq. (A12), overpower series up to quadratic terms, first over small $\xi$, and then, therefore ($r \leq R$, or $x \leq \xi$) over a small variable, $x = r/R_S$. Finally, one obtains from Eq. (A12)

$$a_0(x,q) = (q-2)(4-\xi^2) + 20x^2,$$
$$a_1(x) = 0, \qquad a_2(x) = 4x^2. \qquad (A13)$$

Using these approximate coefficients, one arrives at much simpler Bessel equation:

$$x^2\mathcal{X}''(x) + (q-2)\left(1 - \frac{3}{2}\xi^2\right)\mathcal{X}(x) = 0. \qquad (A14)$$

This equation has the regular solution, Eq. (11).

## Appendix B

## NS macroscopic energies and masses at zero frequency

The total static macroscopic energy, $E$, can be written in terms of the EoS energy density $\mathcal{E}(\rho)$, Eq. (18),

$$E = \int \mathcal{E}(\rho(\mathbf{r}))\mathrm{d}\mathcal{V} \approx E_V + E_S, \qquad (\mathrm{B1})$$

where $\mathrm{d}\mathcal{V} = e^{\lambda/2}\mathrm{d}V,\ \rho(\mathbf{r})$ is the baryon density, $\lambda$ is the inner Schwarzschild parameter, Eq. (3). The integration is fulfilled over the whole spatial volume $dV$ in our first-order perturbation approach over the dimensionless frequency $I/M^2$. In Eq. (B1), $E_V$ is the volume part of the total energy $E$ in the ETF approach,

$$E_V = 2\pi\overline{\rho}R_S^3\left[\arcsin(\xi) - \xi\sqrt{1-\xi^2}\right], \qquad (\mathrm{B2})$$

where $R_S$ is the Schwarzschild radius in Eq. (4), and $\xi$ is given by Eq. (12). For the surface part $E_S$ in Eq. (B1), one obtains

$$E_S \approx \sigma S, \qquad \sigma = \frac{4a\mathcal{J}(R)\overline{\rho}K_G}{135m} = \frac{16}{45}a\kappa\overline{\rho}c^2\mathcal{J}(R), \qquad (\mathrm{B3})$$

where $S = 4\pi R^2$ is the surface area value, $\mathcal{J}(R)$ is the Jacobian $\mathcal{J}(r)$, Eq. (16), at the NS surface, $r = R$. and $\sigma$ is the leading-order tension coefficient [73, 75, 76]. The lepto dermic parameter $a/R$ is related to the inter-particle interaction constant $\mathcal{C}$, Eq. (18), and incompressibility $K_G$, Eq. (19), both depending on the gravity,

$$\frac{a}{R} = \sqrt{\frac{18\mathcal{C}\overline{\rho}}{K_G}}. \tag{B4}$$

In Eq. (B3), we introduced the relative total dimensionless incompressibility $\kappa$ (Ref. [73]), which is a measure of the gravitational field contribution, in addition to the well-known nuclear matter value,

$$\kappa = \frac{K_G}{12mc^2}. \tag{B5}$$

Here, $K_G$ is given by Eq. (19), and $m$ is a probe particle mass. For nuclear physics, one finds $\kappa \approx 2$ $(KG = K \approx 240MeV,\ \rho_0 c^2 = -b_V n_0,\ b_V \approx -16\, MeV,\ n_0 = 0.16\, fm^{-3})$. For a strong NS gravitational field, one may assume $\kappa \gg 1$, and the expression (B3) for the tension coefficient $\sigma$ is accurately motivated within the ESA [75].

## Appendix C

## Macroscopic calculations of NS moments of inertia

For calculations of the statistically averaging $\widetilde{\Theta}_V$, Eq. (22), and correlation $\mathcal{T}_{t\varphi}$, Eq. (23), contributions into Eq. (21) for the MI of a slow rotation taking for a little deformed NS surface, one may introduce the oblate spheroidal coordinate system with the $z$ symmetry axis.

**1. Volume statistically averaging component.** For the volume statistically averaging component $\widetilde{\Theta}_V$ in Eq. (22), one arrives at

$$\widetilde{\Theta}_V = \frac{15\Theta_{sph}}{2}\left(\overline{f}\right)^5 \tilde{\jmath}, \tag{C1}$$

where $\Theta_{sph} = \frac{2}{5}MR^2$ is the spherical rigid body MI, $\overline{f} = f/R = \sqrt{a^2 - b^2}/R \approx \sqrt{2(1 - b/a)}$ is the deformation parameter, $f = \sqrt{a^2 - b^2}$ is the distance between the two spheroid foci, a and b are the oblate spheroid semi-axes, $a > b$. For small deformation $f/R$ up to second order, one obtains the analytical expression for the integral in Eq.(22) over the inner oblate spheroidal coordinates $\{\zeta, \eta, \varphi\}$, $0 \le \zeta \le \zeta_b, -1 \le \eta \le 1, 0 \le \varphi \le 2\pi$, where $\zeta = \zeta_b$ is the NS spheroidal surface,

$$\tilde{\jmath} = \frac{8\zeta_b(1 + \zeta_b^2)^2}{15(2A - 1)^3}\left[2A + \frac{2}{7}\overline{f}\xi^2\left(1 + \frac{5}{4}\zeta_b^2\right) - 1 - \frac{6}{7}\overline{f}^2\xi^2\left(1 + \frac{5}{4}\zeta_b^2\right)\right], \tag{C2}$$

where $\zeta_b = \operatorname{arctanh} Q = \operatorname{arctanh}\left(\frac{b}{a}\right)$, $Q = b/a \approx 1 - f/R < 1$ is the deformation parameter ($a^2 b = R^3$), $\xi$ and $A$ are given by Eqs. (B2) and (4), respectively.

**2. Volume correlation component.** Substituting the inner solution (11) for $X_{in}(r)$ of the $\tau$, Eqs. (7), with (A5) for $Y(\cos\theta)$ into Eq. (23) for the correlation MI component, one finds

$$\mathcal{J}_{t\varphi}^{(V)} = \frac{6c_1 \bar{f}^{7/2}}{\xi^3} \mathcal{I}_{t\varphi}, \tag{C3}$$

Where ξ is given by Eq. (B2),

$$\mathcal{I}_{t\varphi} = \int_0^{\zeta_b}\int_0^1 \frac{\mathrm{d}\zeta \mathrm{d}\eta}{(1+\zeta^2-\eta^2)^{3/2}} \frac{(\zeta^2+\eta^2)(1+\zeta^2)(1-\eta^2)}{\sqrt{1-\bar{f}^2(1+\zeta^2-\eta^2)}\left[A_S - \sqrt{1-\bar{f}^2(1+\zeta^2-\eta^2)/2}\right]^2}$$

$$\times J_p\left(\sqrt{5}\,\bar{f}\sqrt{1+\zeta^2-\eta^2}\right) F\left[\frac{3+\sqrt{9-4q}}{2}, \frac{3-\sqrt{9-4q}}{2}; 2; \frac{1}{2}\left(1+\frac{\zeta\eta}{\sqrt{1+\zeta^2-\eta^2}}\right)\right]. \tag{C4}$$

In order to obtain the analytical approximate results, one can first try to fix the constant index $p(\xi, q)$ of the Bessel function Jp, Eq. (12), and $\alpha(q), \beta(q)$ arguments of the hypergeometric function $F$. Then, we expand the Bessel, Eq. (11), and hypergeometric, Eq. (A5), functions over their arguments for a fixed $p, \alpha$, and $\beta(\gamma = 2)$, Eq. (9), and $A$, Eq. (4). Expanding then the result over the parameter $\bar{f} = f/R$ up to leading terms, one can an analytically take this double integral in terms of the hypergeometric functions: $\mathcal{I}_{t\varphi} = \mathcal{F}(\zeta_b)$, where

$$\begin{aligned}
\mathcal{F}(z) = &\frac{2^{-p}5^{p/2-1}z^{1-p}\left(z\bar{f}R/R_S\right)^p}{21p(1-2A)^2(p^2-4)\Gamma(1+p)} \\
&\times \{-7p(12-4p-3p^2+p^3)(4+\alpha\beta) \\
&\times F\left(\frac{1}{2}, \frac{1-p}{2}, \frac{3}{2}, -z^2\right) \\
&+2p(p^2-4)(4+\alpha\beta)[2(p-3)+7] \\
&\times F\left(\frac{1}{2}, \frac{3-p}{2}, \frac{3}{2}, -z^2\right) \\
&+7z\big[15\alpha\beta(1+z^2)\big(2z^p - 2(1+z^2)^{p/2} \\
&+p(1+z^2)^{p/2} + 2z^2\left(z^p - \left(1+z^2\right)^{p/2}\right)\Big) \\
&+2pz(p^2-4)(4+\alpha\beta)\left[2F\left(\frac{3}{2}, \frac{3-p}{2}, \frac{5}{2}, -z^2\right)\right. \\
&\left.\left.+z^2F\left(\frac{5}{2}, \frac{3-p}{2}, \frac{7}{2}, -z^2\right)\right]\right]\Bigg\}
\end{aligned} \tag{C5}$$

Here, $z = \zeta_b$, $p$ is the Bessel function index, Eq. (12); $\alpha$ and $\beta$ are the coefficients of the hypergeometric functions $F(\alpha, \beta; 2; w)$; see Eq.(9) for $\alpha = \alpha_+$ and $\beta = \alpha_-(\mathcal{F}(z=0)=0)$.

**3. Surface statistical and correlational components.** For the surface parts, $\widetilde{\Theta}_S$ and $\mathcal{T}_{t\varphi}^{(S)}$, of the statistically averaged $\widetilde{\Theta}$, Eq. (22), and cor relational $\mathcal{T}_{t\varphi}$, Eq. (23), one can reduce them to the simple one-dimensional integrals. For the statistical surface part $\widetilde{\Theta}_S$, with the simplest inter-particle interaction $\mathcal{C}$ for small oblate deformations, one finds

$$\tilde{\Theta}_S = \frac{\mathcal{C}}{c^2}\int \mathrm{d}\mathcal{V}\mathcal{J}(r)\varrho^2 e^{-\nu_{\rm in}}(\nabla\rho)^2. \tag{C6}$$

Similarly, for the correlation contribution $\mathcal{T}_{t\varphi}^{(S)}$, one obtains

$$\mathcal{T}_{t\varphi}^{(S)} = \frac{\mathcal{C}}{Mc^2}\int \mathrm{d}\mathcal{V}\mathcal{J}(r)\varrho^2 e^{-\nu_{\rm in}}(\nabla\rho)^2\frac{\tau}{\varrho^2+z^2}. \tag{C7}$$

Here, $\mathcal{J}(r)$ is the same inner Jacobian, Eq. (16), vin is the inner Schwarzschild metric function, Eq. (3), and τ was derived in Appendix A for the inner $r \leq R$ NS region; see Eqs. (11) for $X_{in}(r)$ and (A5) for $Y(\cos\theta)$ for the inner separable solution, Eq. (7). In Eqs. (C6) and (C7), the integral over $\mathrm{d}\mathcal{V}$ is definite as taken over the inner spatial variables accounting for the gravitational shortness effect. From Eqs. (C6) and (C7), following Ref. [75], one arrives a

$$\frac{\tilde{\Theta}_S}{2MR^2/5} = \frac{10\pi}{3}\frac{\sigma R^2\mathcal{J}^2(R)}{Mc^2}, \tag{C8}$$

where $\mathcal{J}(R)$ is the Jacobian $\mathcal{J}(r)$, Eq. (16), at the NS surface $r = R$, and

$$\mathcal{T}_{t\varphi}^{(S)} = \frac{4\pi}{3}\frac{\sigma R^2}{Mc^2}\frac{r_g}{R} = \frac{8\pi\sigma R^2 G}{Rc^4}. \tag{C9}$$

We neglected higher order terms $\sim (a/R)(f/R)$, and used the general ESA method at first order over $a/R$ (see, Refs. [73, 75]) with the help of equation for the derivative of the density $\rho(r)$ over $r$ through the effective surface,

$$\frac{d\rho}{dr} = -\sqrt{\frac{\varepsilon_G(\rho)}{\mathcal{C}}},$$

where $\varepsilon_G(\rho)$ is given by Eq. (19). We used also the variable transformation, $\nabla\rho \approx \partial\rho/\partial r$, and Eq. (B3) for the tension coefficient $\sigma$ within the same accuracy. All surface MI components are proportional to the leptodermic parameter $a/R$ through the surface tension coefficient $\sigma$. As shown in Ref. [73], this coefficient can be calculated analytically; see Eq. (B3).

## REFERENCES

[1] R. C. Tolman, Static solutions of einstein's field equations for spheres of fluid, Phys. Rev. 55, 364 (1939). https://doi.org/10.1103/PhysRev.55.364

[2] R. Tolman, Relativity, Thermodynamics, and Cosmology, Dover Books on Physics (Dover Publications, 1987).

[3] K. Schwarzschild, On the gravitational field of a mass point according to einstein's theory (1999), arXiv:physics/9905030 [physics.hist-ph].

[4] K. Schwarzschild, On the gravitational field of a sphere of incompressible fluid according to einstein's theory (1999), arXiv:physics/9912033 [physics.hist-ph].

[5] L. Landau and E. Lifshitz, The Classical Theory of Fields: Volume 2, Course of theoretical physics (Butterworth-Heinemann, 1975).

[6] S. Shapiro and S. Teukolsky, Black Holes, White Dwarfs and Neutron Stars: The Physics of Compact Objects (John Wiley & Sons, Ltd, 1983). https://doi.org/10.1002/9783527617661

[7] P. Haensel, A. Y. Potekhin, and D. G. Yakovlev, Neutron stars 1: Equation of state and structure, Vol. 326 (Springer, New York, USA, 2007). https://doi.org/10.1007/978-0-387-47301-7

[8] R. X. Adhikari, L. A. Anchordoqui, K. Fang, B. Sathyaprakash, K. Tollefson, et al., Report of the Topical Group on Cosmic Probes of Fundamental Physics for Snowmass 2021, (2022), arXiv:2209.11726 [hep-ph].

[9] J. R. Oppenheimer and G. M. Volkoff, On massive neutron cores, Phys. Rev. 55, 374 (1939). http://doi.org/10.1103/PhysRev.55.374

[10] R. P. Kerr, Gravitational field of a spinning mass as an example of algebraically special metrics, Phys. Rev. Lett. 11, 237 (1963). http://doi.org/10.1103/PhysRevLett.11.237

[11] R. H. Boyer and R. W. Lindquist, Maximal analytic extension of the kerr metric, Journal of Mathematical Physics 8, 265 (1967). https://doi.org/10.1063/1.1705193

[12] P. A. Hogan, An interior kerr solution, Lettere al Nuovo Cimento 16, 33 (1976). https://doi.org/10.1007/BF02746870

[13] P. Collas, Simple tests for proposed interior kerr metrics, Lettere al Nuovo Cimento 21, 68 (1978). https://doi.org/10.1007/BF02745324

[14] S. A. Teukolsky, The kerr metric, Classical and Quantum Gravity 32, 124006 (2015). https://doi.org/10.1088/0264-9381/32/12/124006

[15] D. L. Wiltshire, M. Visser, and S. M. Scott, eds., The Kerr Spacetime: Rotating Black Holes in General Relativity (Cambridge University Press, Cambridge, 2009).

[16] J. B. Hartle, Slowly Rotating Relativistic Stars. 1. Equations of Structure, Astrophys. J. 150, 1005 (1967). https://doi.org/10.1086/149400

[17] J. B. Hartle and K. S. Thorne, Slowly Rotating Relativistic Stars. II. Models for Neutron Stars and Supermassive Stars, Astrophys. J. 153, 807 (1968). https://doi.org/10.1086/149707

[18] T. Regge and J. A. Wheeler, Stability of a Schwarzschild Singularity, Phys. Rev. 108, 1063 (1957). https://doi.org/10.1103/PhysRev.108.1063

[19] K. S. Thorne and A. Campolattaro, Non-Radial Pulsation of General-Relativistic Stellar Mod els. I. Analytic Analysis for L ≥ 2, The Astrophysical Journal 149, 591 (1967). https://doi.org/10.1086/149288

[20] J. L. Friedman, L. Parker, and J. R. Ipser, Rapidly Rotating Neutron Star Models, Astrophys. J. 304, 115 (1986). https://doi.org/10.1086/164149

[21] N. Stergioulas, Rotating stars in relativity, Living Reviews in Relativity 6, 10.12942/lrr-2003-3 (2003). https://doi.org/10.12942/lrr-2003-3

[22] A. Worley, P. G. Krastev, and B.-A. Li, Nuclear Constraints On The Moments Of Inertia Of Neutron Stars, The Astrophysical Journal 685, 390 (2008). https://doi.org/10.1086/589823

[23] J. L. Friedman and N. Stergioulas, Rotating Relativistic Stars (Cambridge University Press, Cambridge, New York, 2013). https://doi.org/10.1017/CBO9780511977596

[24] V. Paschalidis and N. Stergioulas, Rotating Stars in Relativity, Living Rev. Rel. 20, 7 (2017). https://doi.org/10.1007/s41114-017-0008-x

[25] A. Krasi´nski, Ellipsoidal space-times, sources for the kerr metric, Annals of Physics 112, 22 (1978). 10.1016/0003-4916(78)90079-9

[26] Klochkov, D., Suleimanov, V., P¨uhlhofer, G., Yakovlev, D. G., Santangelo, A., and Werner, K., The neutron star in hess31-347: Central compact objects as laboratories to study the equation of state of superdense matter, A&A 573, A53 (2015). https://doi.org/10.1051/0004-6361/201424683

[27] N¨attil¨a, J., Miller, M. C., Steiner, A. W., Kajava, J. J. E., Suleimanov, V. F., and Poutanen, J., Neutron star mass and radius measurements from atmospheric model fits to x-ray burst cooling tail spectra, A&A 608, A31 (2017). https://doi.org/10.1051/0004-6361/201731082

[28] T. E. Riley et al., A NICER View of PSR J0030+0451: Millisecond Pulsar Parameter Estimation, The Astrophysical Journal Letters 887, L21 (2019). https://doi.org/10.3847/2041-8213/ab481c

[29] M. C. Miller, C. Chirenti, and F. K. Lamb, Constraining the equation of state of high-density cold matter using nuclear and astronomical measurements, The Astrophysical Journal 888, 12 (2019). https://doi.org/10.3847/1538-4357/ab4ef9

[30] M. C. Miller et al., The Radius of PSR J0740+6620 From Nicer And XMM-Newton Data, The Astrophysical Journal Letters 918, L28 (2021). https://doi.org/10.3847/2041-8213/ac089b

[31] B. P. Abbott et al., GW170817: Observation of Gravitational Waves from a Binary Neutron Star Inspiral, Phys. Rev. Lett. 119, 161101 (2017). https://doi.org/10.1103/PhysRevLett.119.161101

[32] G. Raaijmakers et al., A Nicer View of PSR J0030+0451: Implications For The Dense Matter Equation of State, The Astrophysical Journal Letters 887, L22 (2019). https://doi.org/10.3847/2041-8213/ab451a

[33] C. D. Capano, I. Tews, S. M. Brown, B. Margalit, S. De, S. Kumar, D. A. Brown, B. Krishnan, and S. Reddy, Stringent constraints on neutron-star radii from multimessenger observations and nuclear theory, Nature Astronomy 4, 625 (2020). https://doi.org/10.1038/s41550-020-1014-6

[34] T. E. Riley et al., A NICER view of the massive pulsar PSR J0740+6620 informed by radio timing and XMM-newton spectroscopy, The Astrophysical Journal Letters 918, L27 (2021). https://doi.org/10.3847/2041-8213/ac0a81

[35] V. Doroshenko, V. Suleimanov, G. P¨uhlhofer, and A. Santangelo, A strangely light neutron star within a supernova remnant, Nature Astronomy 6, 1444 (2022). https://doi.org/10.1038/s41550-022-01800-1

[36] R. Kumar, M. Kumar, V. Thakur, S. Kumar, P. Kumar, A. Sharma, B. K. Agrawal, and S. K. Dhiman, Observational constraint from the heaviest pulsar PSR J0952-0607 on the equation of state of dense matter in relativistic mean field model, Phys. Rev. C 107, 055805 (2023). https://doi.org/10.1103/PhysRevC.107.055805

[37] G. G. L. Nashed, The Effect of *f(R, T)* Modified Gravity on the Mass and Radius of Pulsar HerX1, The Astrophysical Journal 950, 129 (2023), arXiv:2306.10273 [gr-qc]. https://doi.org/10.3847/1538-4357/acd182

[38] D. Sen and A. Guha, Estimating the dark matter halo velocity and surface temperature of some known pulsars due to dark matter capture (2024), arXiv:2402.13795 [hep-ph]. https://arxiv.org/abs/2402.13795

[39] Baglio, M. C., D´Avanzo, P., Mu˜noz-Darias, T., Breton, R. P., and Campana, S., The long term evolution of the X-ray pulsar XTE J1814-338: A receding jet contribution to the quiescent optical emission?, A&A 559, A42 (2013). https://doi.org/10.1051/0004-6361/201322442

[40] Y. Kini et al., Constraining the properties of the thermonuclear burst oscillation source XTE J1814-338 through pulse profile modelling, Monthly Notices of the Royal Astronomical Society 535, 1507 (2024). https://doi.org/10.1093/mnras/stae2398

[41] A. J. Dittmann, M. C. Miller, F. K. Lamb, I. M. Holt, C. Chirenti, M. T. Wolff, S. Bogdanov, S. Guillot, W. C. G. Ho, S. M. Morsink, Z. Arzoumanian, and K. C. Gendreau, A more precise

measurement of the radius of PSR J0740+6620 using updated nicer data, The Astrophysical Journal 974, 295 (2024). https://doi.org/10.3847/1538-4357/ad5f1e

[42] G. Baym, H. A. Bethe, and C. Pethick, Neutron star matter, Nucl. Phys. A 175, 225 (1971). https://doi.org/10.1016/0375-9474(71)90281-8

[43] G. Baym, C. Pethick, and P. Sutherland, The Ground state of matter at high densities: Equation of state and stellar models, Astrophys. J. 170, 299 (1971). https://doi.org/10.1086/151216

[44] R. B. Wiringa, V. Fiks, and A. Fabrocini, Equation of state for dense nucleon matter, Phys. Rev. C 38, 1010 (1988). https://doi.org/10.1103/PhysRevC.38.1010

[45] E. Chabanat, P. Bonche, P. Haensel, J. Meyer, and R. Schaeffer, A Skyrme parametrization from subnuclear to neutron star densities, Nuclear Physics A 627, 710 (1997). https://doi.org/10.1016/S0375-9474(97)00596-4

[46] E. Chabanat, P. Bonche, P. Haensel, J. Meyer, and R. Schaeffer, A skyrme parametrization from subnuclear to neutron star densities part ii. nuclei far from stabilities, Nuclear Physics A 635, 231 (1998). https://doi.org/10.1016/S0375-9474(98)00180-8

[47] A. Akmal, V. R. Pandharipande, and D. G. Ravenhall, Equation of state of nucleon matter and neutron star structure, Phys. Rev. C 58, 1804 (1998). https://doi.org/10.1103/PhysRevC.58.1804

[48] P. Haensel, Neutron star crusts, in Physics of Neutron Star Interiors, Vol. 578 (Springer Berlin Heidelberg, Berlin, Heidelberg, 2001) pp. 127–174.

[49] J. M. Lattimer and M. Prakash, Neutron star structure and the equation of state, The Astro physical Journal 550, 426 (2001). https://doi.org/10.1086/319702

[50] C. J. Horowitz and J. Piekarewicz, Neutron star structure and the neutron radius of $^{208}$Pb, Phys. Rev. Lett. 86, 5647 (2001). https://doi.org/10.1103/PhysRevLett.86.5647

[51] C. J. Horowitz and J. Piekarewicz, Neutron radii of $^{208}$Pb and neutron stars, Phys. Rev. C 64, 062802(R) (2001). https://doi.org/10.1103/PhysRevC.64.062802

[52] I. Sagert, M. Hempel, C. Greiner, and J. Schaffner-Bielich, Compact stars for undergraduates, European Journal of Physics 27, 577 (2006). https://doi.org/10.1088/0143-0807/27/3/012

[53] B.-A. Li, L.-W. Chen, and C. M. Ko, Recent progress and new challenges in isospin physics with heavy-ion reactions, Physics Reports 464, 113 (2008). https://doi.org/10.1016/j.physrep.2008.04.005

[54] N. Chamel and P. Haensel, Physics of neutron star crusts, Living Reviews in Relativity 11, 10 (2008). https://doi.org/10.12942/lrr-2008-10

[55] Z. Arzoumanian, S. Bogdanov, J. Cordes, K. Gendreau, D. Lai, J. Lattimer, B. Link, A. Lom men, C. Miller, P. Ray, R. Rutledge, T. Strohmayer, C. Wilson-Hodge, and K. Wood, X ray timing of

neutron stars, astrophysical probes of extreme physics (2009), arXiv:0902.3264 [astro-ph.HE]. https://arxiv.org/abs/0902.3264

[56] H. Zheng and A. Bonasera, Non-abelian behavior of $\alpha$ bosons in cold symmetric nuclear matter, Phys. Rev. C 83, 057602 (2011). https://doi.org/10.1103/PhysRevC.83.057602

[57] Potekhin, A. Y., Fantina, A. F., Chamel, N., Pearson, J. M., and Goriely, S., Analytical representations of unified equations of state for neutron-star matter, A&A 560, A48 (2013). https://doi.org/10.1051/0004-6361/201321697

[58] A. Bauswein, S. Goriely, and H.-T. Janka, Systematics of dynamical mass ejection, nucle osynthesis, and radioactively powered electromagnetic signals from neutron-star mergers, The Astrophysical Journal 773, 78 (2013). https://doi.org/10.1088/0004-637X/773/1/78

[59] S. Gandolfi, J. Carlson, S. Reddy, A. W. Steiner, and R. B. Wiringa, The equation of state of neutron matter, symmetry energy, and neutron star structure, Eur. Phys. J. A 50, 10 (2014). https://doi.org/10.1140/epja/i2014-14010-5

[60] G. Giuliani, H. Zheng, and A. Bonasera, The many facets of the (non-relativistic) nuclear equation of state, Progress in Particle and Nuclear Physics 76, 116 (2014). https://doi.org/10.1016/j.ppnp.2014.01.003

[61] G. Baym, T. Hatsuda, T. Kojo, P. D. Powell, Y. Song, and T. Takatsuka, From hadrons to quarks in neutron stars: a review, Reports on Progress in Physics 81, 056902 (2018). https://doi.org/10.1088/1361-6633/aaae14

[62] Y. Lim and J. W. Holt, Bayesian modeling of the nuclear equation of state for neutron star tidal deformabilities and GW170817, Eur. Phys. J. A 55, 209 (2019). https://doi.org/10.1140/epja/i2019-12917-9

[63] C. Margaritis, P. S. Koliogiannis, A. Kanakis-Pegios, and C. C. Moustakidis, Crust-core in terface and bulk neutron star properties, Phys. Rev. C 104, 025805 (2021). https://doi.org/10.1103/PhysRevC.104.025805

[64] P. S. Koliogiannis and C. C. Moustakidis, Thermodynamical description of hot, rapidly rotating neutron stars, protoneutron stars, and neutron star merger remnants, The Astrophysical Journal 912, 69 (2021). https://doi.org/10.3847/1538-4357/abe542

[65] B. Sun, S. Bhattiprolu, and J. M. Lattimer, Compiled properties of nucleonic matter and nuclear and neutron star models from nonrelativistic and relativistic interactions, Phys. Rev. C 109, 055801 (2024). https://doi.org/10.1103/PhysRevC.109.055801

[66] N. N. Shchechilin, N. Chamel, and J. M. Pearson, Unified equations of state for cold nonaccreting neutron stars with brussels-montreal functionals. iv. role of the symmetry energy in pasta phases, Phys. Rev. C 108, 025805 (2023). https://doi.org/10.1103/PhysRevC.108.025805

[67] A. F. Fantina and F. Gulminelli, Nuclear physics inputs for dense-matter modelling in neutron stars. the nuclear equation of state, Journal of Physics: Conference Series 2586, 012112 (2023). https://doi.org/10.1088/1742-6596/2586/1/012112

[68] H. D. Thi, A. F. Fantina, and F. Gulminelli, Light clusters in the liquid proto-neutron star inner crust, Eur. Phys. J. A 59, 292 (2023). https://doi.org/10.1140/epja/s10050-023-01199-x

[69] J. M. Lattimer, Constraints on the nuclear symmetry energy from experiments, theory and observations, Journal of Physics: Conference Series 2536, 012009 (2023). https://doi.org/10.1088/1742-6596/2536/1/012009

[70] S. Saklany, N. Pant, and B. Pandey, Three-layered compact star in modified buchdahl-i spatial metric with distinct equations of state, Physics Letters B 845, 138176 (2023). https://doi.org/10.1016/j.physletb.2023.138176

[71] N. Chamel, J. M. Pearson, and N. N. Shchechilin, Role of neutron pairing with density-gradient dependence in the semimicroscopic treatment of the inner crust of neutron stars, Phys. Rev. C 110, 045808 (2024). https://doi.org/10.1103/PhysRevC.110.045808

[72] D. Kobyakov and X. Vi˜nas, Inner crust of neutron stars: Polymorphism and superconductivity in the liquid drop model (2025), arXiv:2411.17303 [nucl-th]. https://doi.org/10.48550/arXiv.2411.17303

[73] A. G. Magner, S. P. Maydanyuk, A. Bonasera, H. Zheng, A. I. Levon, T. M. Depastas, and U. V. Grygoriev, Neutron stars as dense liquid drop at equilibrium within the effective surface approximation, International Journal of Modern Physics E 33, 2450043 (2024). https://doi.org/10.1142/S0218301324500435

[74] A. Magner, S. Maydanyuk, A. Bonasera, H. Zheng, T. Depastas, A. Levon, and U. Grygoriev, Leptodermic corrections to the tov equations and nuclear astrophysics within the effective surface approximation, Nuclear Physics A 1064, 123239 (2025). https://doi.org/10.1016/j.nuclphysa.2025.123239

[75] A. Magner, S. Maydanyuk, A. Bonasera, H. Zheng, T. Depastas, A. Levon, U. Grygoriev, A. Uleiev, Rotating neutron stars within the macroscopic effective-surface approximation, 1 (2025). https://doi.org/10.48550/arXiv.2509.13129

[76] A. A. Uleiev, A. G. Magner, S. P. Maydanyuk, A. Bonasera, H. Zheng, S. N. Fedotkin, A. I. Levon, U. V. Grygoriev, and T. Depastas, Macroscopic approaches to rotating neutron stars, Nuclear Physics and Atomic Energy 27, 5 (2026). https://doi.org/10.15407/jnpae2026.01.005

[77] S. G. Frauendorf and C. Guet, Atomic clusters as a branchof nuclear physics, Annual Review of Nuclear and Particle Science 51, 219 (2001). https://doi.org/10.1146/annurev.nucl.51.101701.132354

[78] M. Brack and R. Bhaduri, Semiclassical Physics, Frontiers in physics (Avalon Publishing, 1997).

[79] V. M. Strutinsky and A. S. Tyapin, Quasistatic drop model of the nucleus as an approximation to the statistical model, Exp. Theor. Phys. (USSR) 18, 664 (1964).

[80] V. M. Strutinsky, A. G. Magner, and M. Brack, The nuclear surface as a collective variable, Zeitschrift f¨ur Physik A 319, 205 (1984). https://doi.org/10.1007/BF01415634

[81] V. M. Strutinsky, A. G. Magner, and V. Y. Denisov, Density distributions in nuclei, Zeitschrift f¨ur Physik A 322, 149 (1985). https://doi.org/10.1007/BF01412028

[82] A. G. Magner, A. I. Sanzhur, and A. M. Gzhebinsky, Asymmetry and spin-orbit effects in binding energy in the effective nuclear surface approximation, International Journal of Modern Physics E 18, 885 (2009). https://doi.org/10.1142/S0218301309013002

[83] J. P. Blocki, A. G. Magner, P. Ring, and A. A. Vlasenko, Nuclear asymmetry energy and isovector stiffness within the effective surface approximation, Phys. Rev. C 87, 044304 (2013). https://doi.org/10.1103/PhysRevC.87.044304

[84] J. P. Blocki, A. G. Magner, and P. Ring, Slope-dependent nuclear-symmetry energy within the effective-surface approximation, Phys. Rev. C 92, 064311 (2015). https://doi.org/10.1103/PhysRevC.92.064311

[85] L. Landau and E. Lifshitz, Fluid Mechanics: Volume 6, Course of theoretical physics, Vol. 6 (Elsevier Science, 1987).

[86] J. Rowlinson and B. Widom, Molecular Theory of Capillarity, Dover books on chemistry (Dover Publications, 1982).

[87] J. S. Rowlinson, Translation of J. D. Van der Waals' "the thermodynamic theory of capillarity under the hypothesis of a continuous variation of density", Journal of Statistical Physics 20, 197 (1979). https://doi.org/10.1007/BF01011513

[88] M. Brack, C. Guet, and H.-B. H˚akansson, Selfconsistent semiclassical description of average nuclear properties—a link between microscopic and macroscopic models, Physics Reports 123, 275 (1985). https://doi.org/10.1016/0370-1573(86)90078-5

[89] D. Vautherin and D. M. Brink, Hartree-Fock calculations with skyrme's interaction. i. spherical nuclei, Phys. Rev. C 5, 626 (1972). https://doi.org/10.1103/PhysRevC.5.626

[90] P. Ring and P. Schuck, The Nuclear Many-Body Problem, Physics and astronomy online library (Springer, 2004).

[91] V. M. Kolomietz, A. I. Sanzhur, S. Shlomo, and S. A. Firin, Equation of state and phase transitions in asymmetric nuclear matter, Phys. Rev. C 64, 024315 (2001). https://doi.org/10.1103/PhysRevC.64.024315

[92] V. M. Kolomietz and A. I. Sanzhur, Equation of state and symmetry energy within the stability valley, European Physical Journal A 38, 345 (2008). https://doi.org/10.1140/epja/i2008-10679-1

[93] V. M. Kolomietz and A. I. Sanzhur, Gibbs-Tolman approach to the curved interface effects in asymmetric nuclei, Phys. Rev. C 88, 044316 (2013). https://doi.org/10.1103/PhysRevC.88.044316

[94] P. Danielewicz and J. Lee, Symmetry energy in nuclear surface, International Journal of Modern Physics E 18, 892 (2009). https://doi.org/10.1142/S0218301309013014

[95] X. Vi˜nas, M. Centelles, X. Roca-Maza, and M. Warda, Density dependence of the symmetry energy from neutron skin thickness in finite nuclei, Eur. Phys. J. A 50, 27 (2014). https://doi.org/10.1140/epja/i2014-14027-8

[96] J. Piekarewicz and M. Centelles, Incompressibility of neutron-rich matter, Phys. Rev. C 79, 054311 (2009). https://doi.org/10.1103/PhysRevC.79.054311

[97] T. Nikˇsi´c, D. Vretenar, and P. Ring, Relativistic nuclear energy density functionals: Mean-field and beyond, Progress in Particle and Nuclear Physics 66, 519 (2011). https://doi.org/10.1016/j.ppnp.2011.01.055

[98] B. T. Reed, F. J. Fattoyev, C. J. Horowitz, and J. Piekarewicz, Implications of PREX-2 on the equation of state of neutron-rich matter, Phys. Rev. Lett. 126, 172503 (2021). https://doi.org/10.1103/PhysRevLett.126.172503